# Communication Bandwidth for Emerging Networks: Trends and Prospects

Sudhir K. Routray, Pallavi Mishra
Department of Telecommunication Engineering
CMR Institute of Technology, Bangalore, India
{sudhirkumar.r, pallavi.m}@cmrit.ac.in

Sutapa Sarkar, Abhishek Javali, Swathi Ramnath
Department of Electronics and Communication Engineering
CMR Institute of Technology, Bangalore, India
{sutapa.s, abhishek.j, swathi.r}@cmrit.ac.in

*Abstract*—Bandwidth is one of the essential resources for communication. Due to the emergence of large number of new services in different types of communications and their value added entities, the demand for bandwidth has gone up more than ever before. The Internet and its allied services are one of the main users of the global bandwidth. With this increasing demand, effective usage of the available bandwidth and the discovery of new bands become very important. In this article, we show current bandwidth usages and their utilities in different application domains. We show the present trends of bandwidth used for global communication by taking the international bandwidth of the core networks in to account. We analyzed the bandwidth trends in optical and wireless communication domains. Emerging services such as Internet of Things and their bandwidth provisioning too have been discussed in this article.

*Keywords—Global bandwidth; global Internet bandwidth; bandwidth scarcity; new bands for emerging networks*

## I. Introduction

Several new networks and services have emerged in the recent years. Networks for the emerging services such as the fifth generation mobile (5G), Internet of Things (IoT) and hybrid optical-wireless access need new spectrum. Because the existing spectrum is either not enough or simply not suitable for these networks. For electronic communication three resources are essential. They are: transmit power, bandwidth and communication infrastructure. Bandwidth is the frequency or group of frequencies used for carrying information from the source to the end customers. Communication infrastructure is the combination of sensors, hardware, middleware, software and actuators (including medium for end-to-end connected communications) used for the electronic transmission and reception. Out of these three resources, bandwidth is very much unique. It is limited and its reuse has to be carefully planned so that the same frequencies must not be used more than once over the same channel. Due to the emerging new services and their value added entities the demand for bandwidth is very much elevated. For smooth and uninterrupted service provisioning without conflicts, new bandwidth has to be discovered or the existing bandwidth has to be reallocated for different services.

Usable bandwidth for communication is a limited resource. It has several constraints and requirements for different applications. Thus spectrum harvesting in emerging services is an important research area [1]. Especially, in the emerging wireless networks and their applications spectrum harvesting is essential as the data rates and other performance metrics demand more spectrum [1]. Spectrum scarcity is found at different instances due to the emergence of new services and applications [2]. It is solved by the engineering communities in different ways such as discovery of new spectrum, reallocation of the existing spectrum and improving the spectral efficiency. Historical bandwidth usage in different types of communication services is found in [3]. Current global bandwidth is mainly provided by the optical fibers in core, regional and metro networks. Global bandwidth survey of TeleGeography provides the broad trends [4]. It shows large growths in the bandwidth in the recent years. The largest share of the global bandwidth is used by the Internet. It is also the main contributor of information collection and dissemination which is normally done through the web portals. The recent trends of websites and the web portals can be found in [5]. Consumer electronics and other strategic growth sectors use a lot of bandwidth in both wireless and wired forms [6]. Emerging wireless communications such as 5G will have new spectrum as the existing spectrum is not enough for the 5G specifications [7]. MM-waves in three different spectrum ranges are the main candidates for 5G though the extended bands in the 6 GHz LTE ranges are also being considered. Similarly, IoT too needs appropriate frequency bands for the flexible applications. In [8], several issues of IoT have been presented. It also includes the frequency bands used for narrowband IoT (NBIoT) according to the recently proposed standards. In [9] – [11], 5G spectrum planning and new spectrum discovery issues are presented. In [12] – [14], optical transport networks and their statistical properties are presented. In [15], current global bandwidth trends are analyzed taking the main bandwidth providers in to account. In [16], current Internet bandwidth usage and its other associated parameters have been analyzed. It also shows the relative positions of different geographical regions as far as the global bandwidth is concerned.

In this article, we present the current trends of bandwidth usage, main bandwidth providers and their future prospective. We present the new bandwidths being discovered and used for emerging applications. Finally, we present that the bandwidth scarcity complications are no more a threat due to the availability of new bands and the proposed reallocation of existing bands.

The rest parts of the article are organized in 4 different sections. In Section II, we present the historical trends of bandwidth usage. In Section III, we show the present trends of global communication. The role of optical and wireless communications and their bandwidths are analyzed. In Section IV, we present the new bands recently been discovered and the reallocation of the existing bands. In Section V, we conclude the article with the main ideas of the article.

## II. BANDWIDTH USAGE TRENDS OF THE RECENT TIMES

Bandwidth usage is being changed every single year. Several factors affect the bandwidth usage and its demand over the years for different applications. In the recent times, the main bandwidth is being devoted for the applications of the Internet and several other communication networks. Global bandwidth for communication is surveyed by TeleGeography [15] and International Telecommunication Union (ITU). Using the information from these two organizations we show the global bandwidth over the years in Fig. 1. By the end of 2015, global bandwidth is estimated to be very close to 300 Tb/s. In [16], information and communication technologies (ICTs) usage facts and statistical data of ITU are presented. It provides global Internet usage statistics around the world. According to [16], net global bandwidth used for the Internet globally is more than 185 TB/s at the beginning of 2016. However, the bandwidth usages are very much unequally distributed across the continents. Europe remains the dominant user of the Global Internet Bandwidth which is close to 80 TB/s followed by Asia Pacific and Americas (North and South America together). On the otherhand Africa is the lowest user of this bandwidth. Likewise, the average Internet bandwidth used by the individuals is the highest (= 131 kB/s) in Europe and the lowest in Africa (= 6 kB/s). It is noteworthy that only 47% of the global population use the Internet, meaning 53% are still outside the ambit of the Internet [16]. Based on the ICD development different countries are provided rankings known as ICT development index (IDI).

IoT usage, mobile cellular subscriptions and the Internet usage follow almost a proportional trend, meaning where the mobile subscriptions are higher, the IoT usage is higher and vice versa. However, in the developed countries the penetrations of IoT and machine type communications are more in comparison to the developing countries. New wireless and cellular communication technologies use new bands of frequencies. The value added services such as IoT and several machine type communications too need new bands.

## III. MAIN COMMUNICATION BANDWIDTHS

Bandwidth despite being the unique resource can be classified for different usages and utilities. The bandwidth used for core networks these days is very much optical. For intercontinental, continental, national, and even regional networks, C-band optical frequencies are very popular. However, the local optical access networks use either L or UL-band frequencies.

### A. Bandwidth for Core, Regional and Metro Networks

These days the core, regional, and metro networks are all optical in nature. Optical fibers are the main carrier of global communication traffic. Core networks are the main traffic carriers and almost all of them are now high speed optical communication links. A list of core optical communication networks can be found in [12] – [14]. All these networks carry 100s of Gb/s data rates. High data rates are sent through the single mode fibers. The signal degradations in optical fibers in the long range emanating from nonlinearities, dispersion, attenuation and other component related factors are compensated using the signal processing based techniques. Optical communication has several bands of usable frequencies. However for the long range communication, C-band is widely used. It is normally in the range of 1530 nm to 1570 nm. WDM is the popular choice of multiplexing in the core networks. However, in the recent days, OFDM is also preferred for optical communications as it provides elastic optical networking. Meaning the bandwidth of a specific channel can be increased or decreased dynamically using OFDM.

Regional and metro networks are the branches of the core networks. Due to the large demands for data rates these networks too carry large data rates. Like the core networks these networks also have high capacities. The requirements of these networks and the techniques applied are very much similar to the core networks. These networks also use C-band frequencies.

### B. Bandwidth for Access Networks

Access area networks are dominated by the wireless service networks. Mobile and other hand handheld devices are the main interfaces for the delivery of communication contents in different formats. In different mobile generations separate bands are used for communication. Right now, in most parts of the world 2G, 3G, 4G and advanced versions such as 4.5G coexist. Therefore, the spectrum used for these services should not be overlapped. Microwaves are the popular choices for wireless access networks.

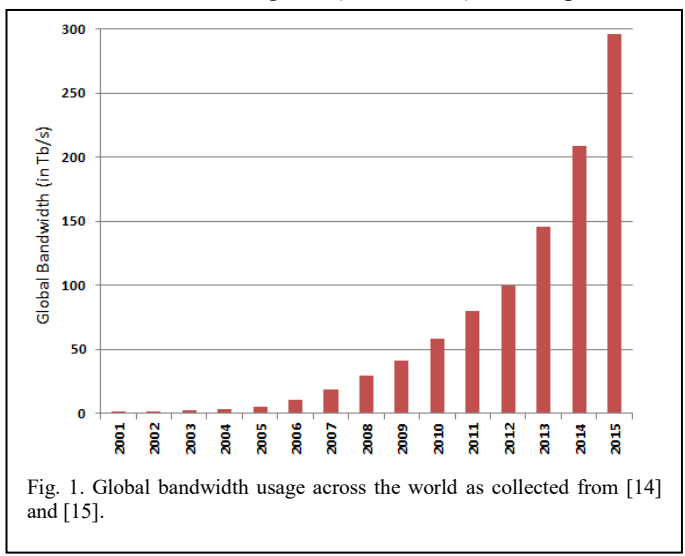

Fig. 1. Global bandwidth usage across the world as collected from [14] and [15].

In the wired domain, different frequencies are used. Broadband communications are the main bandwidth consumption agents in the access areas. Broadband services are very much integrated meaning several services are combined and sent over the same wired medium. At the receiving end they are separated out using an appropriate modem. Digital subscriber lines (DSLs) use different frequencies than the optical fibers. Passive optical networks (PONs) which lead in the optical access areas provide the fastest broadband services. Normally, PONs use UL band of the optical spectrum which is centered around 1625 nm. Of course in some countries the 1300 nm bands too are used for PONs. However, this band is not as popular as the UL band for optical access networks.

## IV. BANDWIDTH FOR EMERGING SERIVES

Bandwidth is scarce and existing bandwidth has to be utilized efficiently. Every year new applications are emerging in communications in various forms. Since the 1980s, mobile communications are evolving and new generations arrive every decade. 5G is the new trend currently which will hit the market around 2020. 5G will will have several value added services. For these services, either a part of 5G bandwidth will be used or new bandwidth will be discovered.

### A. Discovery of New Spectrum

Discovery of new spectrum for communication purpose is the direct solution to spectrum scarcity. Here, discovery of new spectrum indicates the use of specific bands of spectrum for communication purpose for the first time. In fact, these spectra are known to the scientific community, but, have not been utilized for communication purpose before. For instance, use of 6 GHz band for WiFi was started when the traditional 2.54 GHz range frequencies became insufficient. Likewise, several new bands of frequencies have been discovered for communication purpose in radio, microwave and beyond microwave ranges. In this subsection, we go through the new spectrum discovery and its related aspects in communication.

*1) MM Waves for Cellular Comunication:* Millimiter (MM) waves are part of the microwaves. Their wavelengths are in the MM range. Normally, they are from 30 GHz to 300 GHz. These waves need smaller antennas for their reception. Their attenuation and absorption characteristics too are different from the rest parts of microwave.

*2) Terahertz Range Frequencies:* Frequencies beyond 300 GHz till 3 THz are known as terahertz frequencies. These frequencies were not studied properly by the scientific community till the 1990s. Now, it is found that they have several elegant proporties (in attenuation, absorption and diffraction) which are suitable for communication in the access areas.

### B. Reallocation of Spectrum

Reallocation of spectrum does not introduce new spectrum to solve the spectrum scarcity. Rather, it reorganizes and reallocates the existing spectrum among the competing communication services.

*1) Reallocation of Radio and Microwave Spectrum:* Reallocation of radio waves and microwaves have been carried out several times in the past. It happens in almost all the developed countries in which there are several commuication services. FCC reallocates the radio and microwave spectrum in every one to two decades since the 1960s. The recent reallocation had happened in 2004. In reallocation of spectrum the existing spectrum is divided among all the competing parties based on some logical concept. For the commercial purposes, spectrum is auctioned and the bidders compete for their preferable spectrum. The governments do the auction as they are the legal owner of the usable spectrum in their own countries.

*2) Reallocation of Extended Bands:* Extension of existing bands is possible if appropriate technology is available for the adaptation.

*3) Reallocation of the Spectrum Holes and Guardbands:* Spectrum holes are the unused frquency bands between allocated bands. Guardbands are the frequency holes or unused freuency bands which separate the used bands from the adjacent bands.

*4) Dynamic Allocation of Spectrum Holes:* Dynamic spectrum holes are allocated but unused freuency bands. Dynamic allocation of allocated spectrum holes is possible using spectrum sensing technologies. All the allocated bands are not utilized all the time. The licence holder of the spectrum is the primary user. If another user utilizes the allocated band then this new user (who does not posses the lisence) is known as secondary user. The unused spectrum can be used for some high bandwidth services using smart methods such as cognitive radio applications. In cognitive radio, the dynamic spectrum holes are identified and used by the seconday users.

### C. Bandwidth for 5G

5G is now envisioned as the most advanced version of mobile communications whose deployment testing and performance checking will take place during 2017 – 2019 [9]-[11]. It is expected to be rolled out in 2020. Large scale commercial deployment will happen after 2020. 5G would most probably use the mm waves. Three different bands have been proposed for 5G: 27 – 30 GHz; 57 – 60 GHz and 72 – 75 GHz [11]. However, the LTE development groups are also looking for the unlicensed bands around 6 GHz. Recently several companies including ZTE, China Unicom, KDDI, Docomo and Samsung have tested 5G in the sub 6 GHz bands. However, for wide band operations for a dense environment, the mm-waves are preferred over the sub 6 GHz bands.

### D. Bandwidth for IoT

IoT is an emerging service for connecting all sensor based devices. Billions of devices with heterogeneous performance requirements will be connected through IoT. Classification of IoT is possible based on the performances and system requirements. 5G IoT will be associated with Fifth Generation mobile communications and it will use 5G spectrum. Its spectrum for individual users will be wider than the currently available IoTs. At present, sub 6 GHz bands as well as the mm-wave bands are the main candidates for 5G. However, narrowband IoT (NBIoT) is already in use. It has a large market and its growth perspectives too are very attractive. NBIoT uses three different types of bands: standalone bands; guard bands; and in-bands. Standalone bands utilize the unlicensed frequency bands around the GSM and UMTS bands. Right now, 700MHz – 800 MHz bands are popular for NBIoT in several countries in Europe. Guard bands are the band gaps between the allocated frequency bands. These bands are not used by GSM or its LTE hierarchies. That is why the operators often assume it as the wastage of bandwidth. But, now its new applications in NBIoT are making it useful. In-band NBIoT uses the some of the existing GSM or LTE frequency slots. It is normally allocated using some frequency hopping algorithm so that the LTE (or GSM) and NBIoT services do not collide in the same band. In Fig. 2, we show the NBIoT frequency allocations.

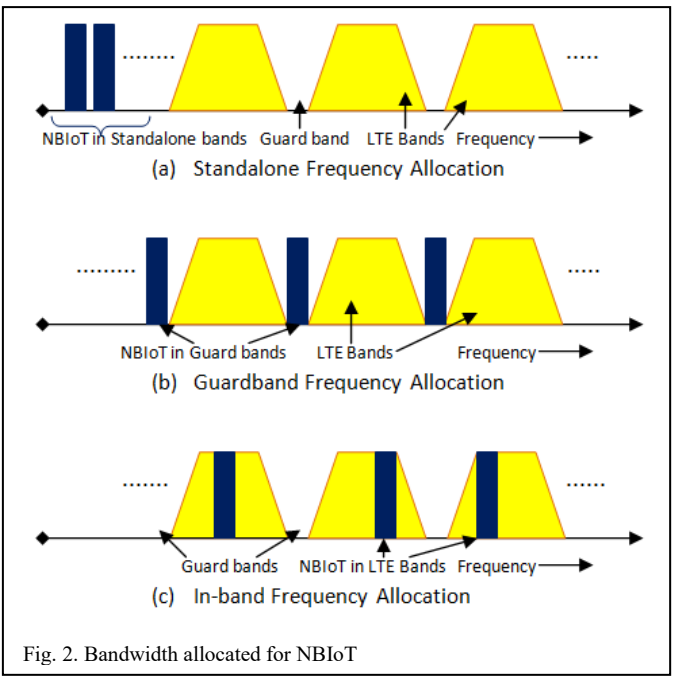

Fig. 2. Bandwidth allocated for NBIoT

### V. CONCLUSIONS

Bandwidth is a very unique resource for communication. It cannot be reused like other resources, rather meticulous planning is required for its repeated usage (such as space division multiplexing). Unlike some other resources it also cannot be replicated whenever required. In this article, we present the past and current trends of communications from the mid 18[th] century till the current years. We also extend the analysis for the future. We project the future prospects of bandwidth usage by taking the current research directions in focus. We presented the current state of the art of global communication bandwidth scenarios and its main future directions. We found that the dangers of bandwidth scarcity have been eased due to the discovery of new spectrum. Effective usage of spectrum has been improved significantly. New spectrum discovered in the previously unused bands is very much promising. For 5G mm-waves are very suitable, though its field trials are still underway.